\documentclass[5p,twocolumn,number]{elsarticle}

\usepackage{epsfig}

\begin{document}

\begin{frontmatter}

\title{Search for Solar Axions Produced by Primakoff Conversion Using Resonant Absorption by $^{169}$Tm Nuclei}

\author[pnpi]{A.V.~Derbin}
 \ead{derbin@pnpi.spb.ru}

\author[pnpi]{S.V.~Bakhlanov}

\author[pnpi]{A.~I.~Egorov}

\author[pnpi]{I.~A.~Mitropol'sky}

\author[pnpi]{V.~N.~Muratova}

\author[pnpi]{D.~A.~Semenov}

\author[pnpi]{E.~V.~Unzhakov}

\address[pnpi]{St.Petersburg Nuclear Physics Institute, Gatchina,
Russia 188300}

\begin{abstract}
The search for resonant absorption of the Primakoff solar axions by
$^{169}$Tm nuclei have been performed. Such an absorption should
lead to the excitation of low-lying nuclear energy level:
$A+^{169}$Tm $\rightarrow ^{169}$Tm$^*$ $\rightarrow ^{169}$Tm $+
\gamma$ (8.41 keV). The Si(Li) detector and $^{169}$Tm target placed
inside the low-background setup were used for that purpose. As a
result, a new restriction on the axion-photon coupling and axion
mass was obtained: $g_{A\gamma}(\mbox{GeV}^{-1})\cdot
m_A(eV)\leq1.36\cdot10^{-5}$ (90\% c.l.). In model of hadronic axion
this restriction corresponds to the upper limit on axion mass -
$m_A\leq$ 191 eV for 90\% c.l.
\end{abstract}


\end{frontmatter}

\section{Introduction}
The appearance of an axion in theory is connected with the problem
of CP-violation in strong interactions. In order to solve this
puzzle Peccei and Quinn \cite{Pec77} proposed the concept of the new
chiral symmetry $U(1)$. The spontaneous breaking of this symmetry at
the energy $f_{A}$ allows one to compensate CP-violating term of the
QCD Lagrangian completely. Weinberg \cite{Wei78} and Wilczek
\cite{Wil78} showed that the introduced model should lead to the
occurrence of a new pseudoscalar particle. The axion mass $m_{A}$
appears to be inversely proportional to the $f_{A}$ value, as well
as the effective axion coupling constants with photons
($g_{A\gamma}$), leptons ($g_{Ae}$), and hadrons ($g_{AN}$). The
model of "standard" or PQWW-axion, where the value of $f_{A}$ was
fixed at the electroweak scale ($f_{A}\approx\sqrt{2}G_F)^{-1/2}$
has been excluded by the series of experiments with radioactive
sources, reactors and accelerators.

Two types of the "invisible" axion models retained the axion in the
form required for the solution of CP-violation problem, while
suppressing its interaction with matter, since the scale of symmetry
breaking $f_{A}$ appears to be arbitrary in these models and can be
extended up to the Planck mass $m_{P} \approx 10^{19} GeV$. These
are models of "hadronic" or KSVZ axion \cite{Kim79,Shi80} and the
GUT or DFSZ axion \cite{Zhi80,Din81}. The axion mass $m_{A}$ (in
$eV$ units) in both models is given in terms of $\pi^0$-properties:
\begin{equation}\label{ma}
m_A=\frac{f_{\pi}m_{\pi}}{f_A}(\frac{z}{(1+z+w)(1+z)})^{1/2}\approx
\frac{6.0\cdot 10^6}{f_A(GeV)}
\end{equation}
where $f_\pi\cong$ 93 MeV is the pion decay constant; $z = m_u/m_d
\cong 0.56$ and $w = m_u/m_s \cong0.029$ are quark-mass
ratios\footnote{We follow the generally accepted units: GeV for
$f_A$ and $g_{A\gamma}$ and eV for $m_a$.}.

The restrictions on the axion mass appear as a result of the
restrictions on the coupling constants $g_{A\gamma}$, $g_{Ae}$ and
$g_{AN}$, which are significantly model dependent. The hadronic
axion does not interact with leptons and ordinary quarks at the tree
level, which results in strong suppression of $g_{Ae}$ coupling
through radiatively induced coupling. Moreover, in some models
axion-photon coupling may significantly differ from the original
DFSZ or KVSZ $g_{A\gamma}$ couplings . The coupling constant
$g_{A\gamma}$ for the "invisible" axion models is equal to:
\begin{equation}\label{gagamma}
    g_{A\gamma}=\frac{\alpha}{2\pi f_A}\left(\frac{E}{N}-\frac{2(4+z)}{3(1+z)}\right)=\frac{\alpha}{2\pi f_A}C_{A\gamma}
\end{equation}
where $\alpha\approx1/137$ is a fine structure constant, $E/N$ is
the ratio of the electromagnetic and colour anomalies, a model
dependent parameter is of the order of unity. $E/N=8/3$ for the
DFSZ-axion ($C_{A\gamma\gamma}$=0.74) and $E/N=0$
($C_{A\gamma\gamma}$=-1.92) in the original KSVZ-axion model. The
value of the second term inside brackets is 1.95$\pm$0.08 and
axion-photon coupling may be reduced by a factor less than 10$^{-2}$
in axion models in which $E/N$ is close to 2 \cite{Kap85}.

If axions do exist, then the Sun should be an intense source of these particles. Axions can be efficiently produced in the Sun
by the Primakoff conversion of photons in the electric field of the plasma. The resulting axion flux depends on $g_{A\gamma}^2$
and can be detected by inverse Primakoff conversion of axions to photons in laboratory magnetic fields \cite{Sik83} -
\cite{Ari09} or by the coherent conversion to photons in the crystal detectors \cite{Avi99}-\cite{Bru08}. The expected count
rate of photons depends on the axion-photon coupling as $g_{A\gamma}^4$.

In this letter we present the results of the search for solar axions
using the other reaction - the resonant absorption by nuclear
target. The $\gamma$-rays and conversion electrons produced by the
de-excitation of the nuclear level can be registered. The detection
probability of the axions is determined by the product
$g_{A\gamma}^2$$\cdot$$g_{AN}^2$ which is preferable for small
$g_{A\gamma}$ values.

There are two other possible mechanisms of axion production in the
Sun: the reactions of solar cycle and the excitation of the
low-lying energy levels of some nuclei by the high solar
temperature. The attempts to detect the resonant absorption of
quasi-monochromatic axions emitted in nuclear magnetic transitions
were performed in \cite{Mor95}-\cite{Bel08}. 
Astrophysical and cosmological data provide restrictive constraints
on axion mass so that acceptable $m_a$ values belong to the
$10^{-4}\div10^{-2}$ eV region \cite{Raf08,PDG08}. The results of
laboratory searches for the axion as well as the astrophysical axion
bounds one can find in \cite{PDG08,Raf06}

\section{The rate of solar axions absorption by $^{169}$Tm nucleus}
\begin{figure}
\includegraphics[width=9cm,height=10.5cm]{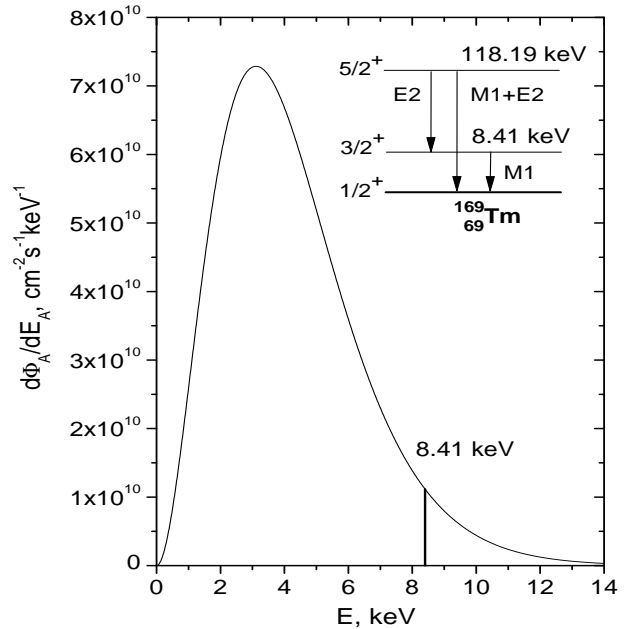}
\caption {Energy spectrum of the axions produced by Primakoff effect
at the Sun (for $g_{A\gamma}=10^{-10}\mbox{ GeV}$). The level scheme
of $^{169}$Tm nucleus is shown in the inset.}\label{fig1}
\end{figure}

The energy spectrum of solar axions produced by Primakoff effect is parameterized by the following expression
\cite{Bib89,Cre98,Zio05}:
\begin{equation}\label{difflux}
    \frac{d\Phi_A}{dE_A}=(g_{A\gamma})^2\cdot 3.82\cdot10^{30}\frac{(E_A)^3}{\exp(E_A/1.103)-1},
\end{equation}
where the value of flux is given in (cm$^{-2}$s$^{-1}$keV$^{-1}$)
units, value of $E_A$ energy is given in keV units and the value of
$g_{A\gamma}$ is given in GeV$^{-1}$ units. The axion spectrum
calculated in assumption that $g_{A\gamma}=10^{-10}$ GeV$^{-1}$ is
given in Fig.\ref{fig1}.
The average energy of axions is $\approx4$ keV, but their flux
becomes negligibly small at the energies above 15 keV.

As a pseudoscalar particle, the axion should be subject to resonant
absorption and emission in the nuclear transitions of a magnetic
type. For our experiment we have chosen the $^{169}$Tm nucleus as a
target. The energy of the first nuclear level (3/2$^+$) is equal to
8.41 keV, the axion flux at this energy is only 7 times less than
the maximum level (fig.\ref{fig1}). The 8.41 keV nuclear level
discharges through $M1$-type transition with $E2$-transition
admixture value of $\delta^2$=0.11\%. Considering the electron
conversion ratio $e/\gamma=263$ \cite{NDS08} one obtains the
relative probability of $\gamma$-ray emission
$\eta=1/(1+e/\gamma)=3.79\cdot10^{-3}$.

The cross-section for the resonant absorption of the axions with the
energy $E_A$ is given by the expression that is similar to the one
for $\gamma$-ray resonant absorption, but the ratio of the nuclear
transition probability with the emission of an axion $(\omega_{A})$
to the probability of magnetic type transition $(\omega_{\gamma})$
have to be taken into account. The rate of solar axion absorption by
$^{169}$Tm nucleus will be
\begin{equation}\label{rate}
    R_A = \pi\sigma_{0\gamma}\Gamma
    \frac{d\Phi_A}{dE_A}(E_A=8.4)\left(\frac{\omega_A}{\omega_\gamma}\right),
\end{equation}
where  $\sigma_{0\gamma}$ is a maximum cross-section of $\gamma$-ray
absorption. The experimentally derived value of $\sigma_{0\gamma}$
for $^{169}$Tm nucleus is equal to
$\sigma_{0\gamma}=2.56\cdot10^{-19}$ cm$^{2}$. A lifetime of the
$^{169}$Tm first excited level is $\tau=5.89$ ns, thus the width of
the energy level $\Gamma=1.13\cdot10^{-7}$ eV.

The $\omega_A/\omega_\gamma$ ratio calculated in the long-wave
approximation, has the following view \cite{Don78,Avi88}:
\begin{equation}\label{omegaaomegag}
    \frac{\omega_A}{\omega_\gamma}=\frac{1}{2\pi\alpha}\frac{1}{1+\delta^{2}}\left[\frac{g^0_{AN}\beta+g^3_{AN}}{(\mu_0-0.5)\beta+\mu_3-\eta}\right]^2\left(\frac{p_A}{p_\gamma}\right)^3.
\end{equation}
Here, $p_\gamma$ and $p_A$ are the photon and axion momenta,
respectively, $\mu_0=\mu_p+\mu_n\approx0.88$ and
$\mu_3=\mu_p-\mu_n\approx4.71$ are isoscalar and isovector nuclear
magnetic momenta, $\beta$ and $\eta$ are parameters depending on the
particular nuclear matrix elements \cite{Avi88,Hax91}. In case of
the $^{169}$Tm nucleus, which has the odd number of nucleons and an
unpaired proton, in the one-particle approximation the values of
$\beta$ and $\eta$ can be estimated as $\beta\approx1.0$ and
$\eta\approx0.5$. For the given parameters the branching ratio can
be rewritten as:
\begin{equation} \label{rat}
\frac{\omega_{A}}{\omega_{\gamma}}=1.03(g_{AN}^{0}+g_{AN}^{3})^2(p_A/p_{\gamma})^3.
\end{equation}

In the KSVZ axion model the dimensionless isoscalar and isovector
coupling constants $g_{AN}^{0}$ and $g_{AN}^{3}$  are related to
$f_A$ by expressions \cite{Kap85,Sre85}:
\begin{equation}\label{g0g3}
g_{AN}^{0}=-\frac{m_N}{6f_A}[2S+(3F-D)\frac{1+z-2w}{1+z+w}]
\end{equation}
and
\begin{equation}
g_{AN}^{3}=-\frac{m_N}{2f_A}[(D+F)\frac{1-z}{1+z+w}]
\end{equation}
where $m_N\approx939$ MeV is the nucleon mass. The axial-coupling
parameters $F$ and $D$ are obtained from hyperon semi-leptonic
decays with a high degree of precision: $F=0.462$ $\pm$ 0.011, $D$=
0.808 $\pm$ 0.006 \cite{Mat05}. The parameter $S$ characterizing the
flavor singlet coupling still remains a poorly constrained one. Its
value varies from $S=0.68$ in the naive quark model down to
$S=-0.09$ which is given on the basis of the EMC collaboration
measurements \cite{May89}. The more stringent boundaries $(0.37\leq
S\leq0.53)$ and $(0.15\leq S\leq0.5)$ were found in \cite{Alt97} and
\cite{Ada97}, accordingly. As a result the value of the sum
($g_{AN}^0+g_{AN}^3$) is determined within a factor of two, but the
ratio $\omega_A/\omega_{\gamma}$ does not vanish for any value of
parameter S, in the case of the transition caused predominantly by
the proton ($\beta\geq$0). We will use $S=0.5$ as reference when we
calculate an axion flux for KSVZ axion model.

The values of $g_{AN}^{0}$ and $g_{AN}^{3}$ for the DFSZ axion
depend on an additional unknown parameter $\cos^2\beta$ which is
defined by the ratio of the Higgs vacuum expectation values. In case
of $^{169}$Tm $M1$-transition the value of
$(\omega_{A}/\omega_{\gamma})^{DFSZ}$ ratio lies within the interval
$\sim$ (0.1$\div$2.0)$(\omega_{A}/\omega_{\gamma})^{KSVZ}$. The
lower and upper bounds of this interval are defined by values
$\cos\beta=0, 1$ respectively.

In accordance with (\ref{difflux}) and (\ref{rat}), the rate of
axion absorption by $^{169}$Tm nucleus (\ref{rate}) dependent only
on the coupling constants is (the model-independent view):
\begin{equation}\label{rategamag0g3}
    R_A=104\cdot g_{A\gamma}^2(g_{AN}^{0}+g_{AN}^{3})^2(p_A/p_\gamma)^3.
\end{equation}
Using the relations between $g_{AN}^0$, $g_{AN}^3$ and axion mass
given by KVSZ model (\ref{g0g3}), the absorption rate can be
presented as a function of $g_{A\gamma}$ and axion mass $m_A$:
\begin{equation}\label{rategama}
    R_A=4.80\cdot10^{-13}g_{A\gamma}^2m_A^2(p_A/p_\gamma)^3.
\end{equation}
At last, one can use the connection between $g_{A\gamma}$ and $m_A$
given by expressions (\ref{ma}) and (\ref{gagamma}), thus the $R_A$
value appears to be proportional to $m_A^4$:
\begin{equation}\label{ratenumeric}
    R_A=6.64\cdot10^{-32}m_A^4(p_A/p_\gamma)^3.
\end{equation}

The amount of observed $\gamma$-rays that follow the axion
absorption depends on the number of target nuclei, measurement time
and detector efficiency, while the probability of 8.4 keV peak
observation is determined by the background level of the
experimental setup.

\section{Experimental setup}
In order to register 8.41 keV $\gamma$-rays we used the planar
Si(Li) detector  with the diameter of the sensitive region $d
\approx 17$ mm and 2.5 mm thick.  The detector was mounted inside a
vertical vacuum cryostat $\approx 3$ mm below the 20 $\mu$m thick
beryllium window. The Tm$_2$O$_3$ target consisting of 306 mg of
$^{169}$Tm with the diameter of 25 mm ($\rho$=31 mg/cm$^2$) was
placed directly on the beryllium window. The tantalum collimator of
8-mm radius was situated between the beryllium window and detector
surface to exclude events at the boundary of the sensitive volume.

The cryostat was enclosed within the concentric copper (\O29 cm) and
lead  (\O34 cm) shells, which provided the shielding against
external radioactivity. The setup was located on the ground surface
and was assembled of five $50\times50\times12$ cm$^3$ plastic
scintillators against the cosmic rays and fast neutrons. The rate of
50 $\mu$s veto signals was 600 counts/s, that leads to $\approx$3\%
dead time. The spectrum of the Si(Li) signals obtained in the
coincidence with veto signals allows to check the probability of
excitation of the 8.41 keV level by the nuclear active component and
cosmic ray muons.

The data-acquisition system of the Si(Li) detector was organized in
the following way: the output of the Si(Li) detector was fed to the
charge-sensitive preamplifier and then to two separate shaping
amplifiers with different gain ratios, which makes it possible to
collect spectra from both lower $(0\div\mbox{60 keV})$ and higher
$(0\div500\mbox{ keV})$ energy regions. This feature allowed us to
monitor the background level over a wide range of natural
radioactivity and, in particular, to detect the $\gamma$-rays
corresponding to the de-exitation of the second exited state of
$^{169}$Tm (118 keV). In such a way we can evaluate the fast neutron
flux which can excite 8.41 keV level. The signal from each amplifier
was received by the individual 4096-channel ADC. Taking into account
the spectra measured in coincidence with the veto system for each
amplifier, four 4096-channel spectra were recorded.

The energy scale was defined using standard calibration sources of
$^{55}$Fe, $^{57}$Co and $^{241}$Am. The prompt energy resolution of
the detector determined by the 14.4 keV $\gamma$-line from $^{57}$Co
turned out to be $\sigma$=FWHM/2.35=120 eV. The high energy
resolution and accurate knowledge of the energy scale is very
important because the energies of the characteristic X-rays of
thulium are very close to 8.41 keV. The most intense L-lines in the
case of vacancy on K-shell possesses the following energies and
intensities: 7.18 keV (8.1\%, L$_{\alpha 1}$), 8.10 keV (5.2\%,
L$_{\beta 1}$) and 8.47 keV (1.6 \%, L$_{\beta 2}$).

The sensitive volume and the area of Si(Li) detector were measured
using the X-rays from $^{241}$Am and $^{55}$Tm sources. The
self-absorption of 8.41 keV $\gamma$-rays by the target was found
via a detailed M-C simulation. The results of simulation were tested
with $^{241}$Am source placed behind the Tm$_2$O$_3$ target. The
overall detection efficiency for 8.41 keV $\gamma$ is estimated to
be (1.28$\pm$0.06)\%.

\section{Results}
\begin{figure}
\includegraphics[width=9cm,height=10.5cm]{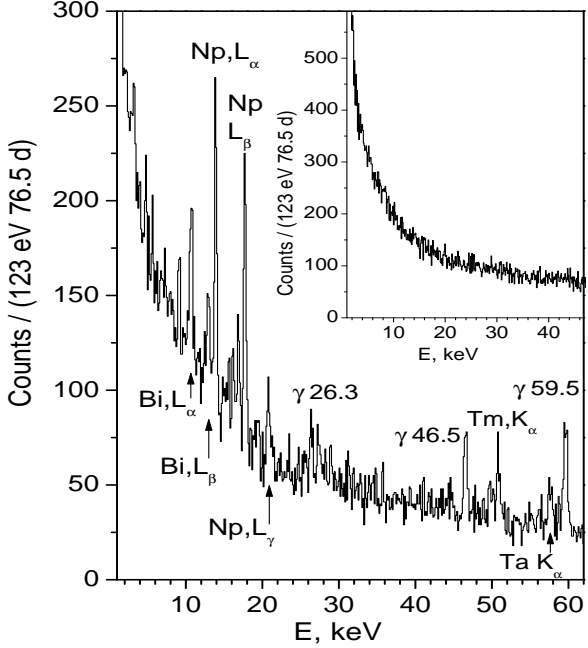}
\caption {The Si(Li)-detector spectra measured during 76.5 days: in
the anticoincidence (main) and in coincidence (inset) with signal of
veto system. } \label{fig2}
\end{figure}

The measurements were carried out during 76.5 days of live time by
brief 2-hour runs in order to monitor the time stability of the
data-acquisition system. The obtained energy spectra of the Si(Li)
detector in the range $(1\div62)$ keV are shown in Fig.\ref{fig2}.
While the spectrum measured in the coincidence with signals of the
veto scintillators does not contain any peculiarity, one can clearly
identify several peaks in the spectrum of uncorrelated events.

Two intense peaks with the energies 13.95 keV  and 17.8 keV  are L
X-rays of neptunium that is produced in the $^{241}$Am
$\alpha$-decay: $^{241}\mbox{Am}\rightarrow^{237}\mbox{Np}$. The
13.9 keV peak consists of two lines with the energies 13.946 keV
(13\%, L$_{\alpha1}$) and 13.761 (1.4\%, L$_{\alpha2}$), the 17.8
keV peak is more complex one formed by L$_{\beta1-5}$ lines. The
less intense peaks with energies of 10.84 keV and 13.2 keV
(L$_{\alpha1}$ and L$_{\beta(1-5)}$ of Bi, respectively) are present
due to $^{210}$Pb$\rightarrow^{210}$Bi $\beta$-decay in $^{238}$U
series. The well known 59.54 keV $^{241}$Am and 46.54 keV $^{210}$Pb
$\gamma$-lines, as well as 50.7 keV K$_{\alpha1}$ of Tm and 57.5 keV
K$_{\alpha1}$ of Ta X-rays, were observed in the high energy part of
the spectrum. The 13.9 keV peak together with 46.54 keV and 59.54
keV $\gamma$-peaks were used to find the final energy scale and the
energy resolution $\sigma (E)$ of the detector. Since no special
electronic stabilization was used, the energy resolution determined
for the 13.9 keV peak(L$_{\alpha1}$ of Np with small admixture of
L$_{\alpha2}$) spread up to $\sigma\cong130$ eV during the long-time
measurements.

\begin{figure}
\includegraphics[width=9cm,height=10.5cm]{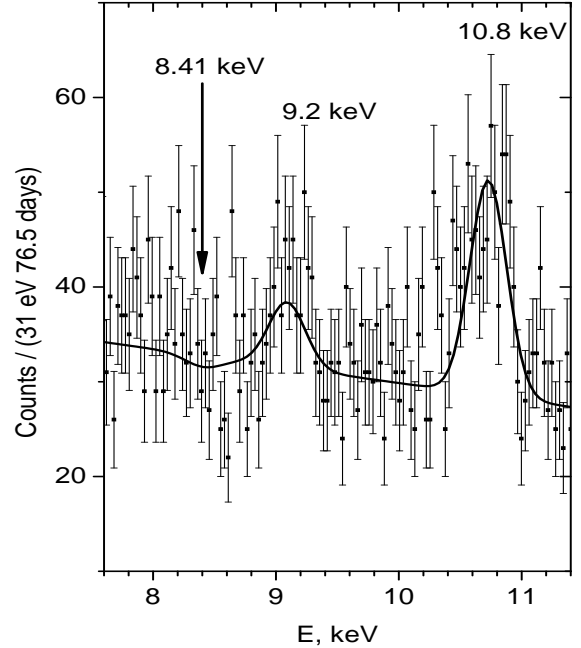}
\caption {Energy spectrum measured with the Si(Li)-detector in the
region 7.6-11.4 keV. The fit is shown by solid line.} \label{fig3}
\end{figure}

Fig.\ref{fig3} shows the detailed energy spectrum within the
(7.6$\div$11.4) keV interval, where the "axion peak" was expected.
There is no pronounced peak at 8.41 keV. In order to determine the
intensity of the 8.41 keV peak we used the maximum likelihood
method. The likelihood function was determined as a sum of three
Gaussians and the linear background assuming that the number of
counts in each channel had normal distribution. The first Gaussian
represents the known characteristic L$_{\alpha1}$ X-rays of Bi
($E_1\cong$ 10.8 keV, the second Gaussian describes the shape of the
peak with the energy $E_2\cong$ 9.2 keV, that we explain by
L$_{\alpha1}$ x-rays of Ir (L$_{\beta1}$ 10.72 keV x-rays of Ir
spread 10.8 keV peak) and the third Gaussian ($S_A$) stands for the
expected 8.41 keV axion peak:
\begin{equation}\label{likelihood}
N(E)=a+b\cdot
E+\frac{1}{\sqrt{2\pi}\sigma}\sum^{3}_{i=1}S_{i}\exp\left[-\frac{(E_{i}-E)^{2}}{2\sigma^{2}}\right].
\end{equation}
The energy resolution ($\sigma$) and the position of 8.41 peak was
fixed during the fitting, while the positions ($E_1,E_2$) and
intensities ($S_1,S_2$, $S_A$) and background coefficients ($a$,$b$)
were independent free parameters.

The fitting result is given in Fig.\ref{fig3} (solid line).
\begin{figure}
\includegraphics[width=9cm,height=10cm]{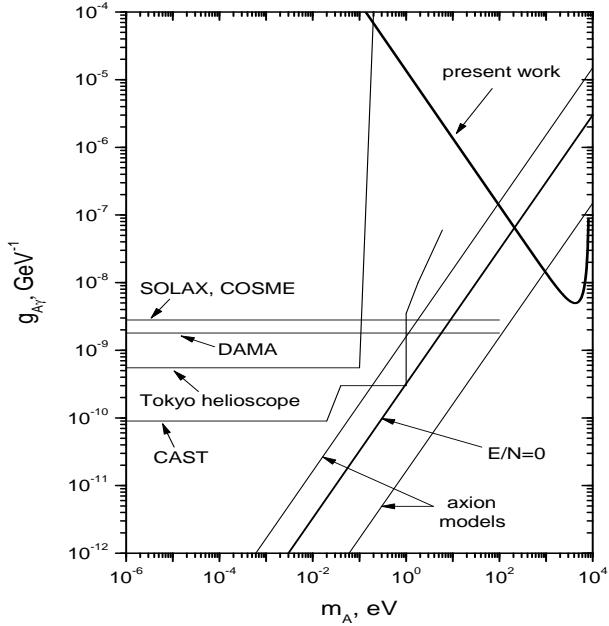}
\caption {The limits on $g_{A\gamma}$ coupling constant obtained by Solax \cite{Avi99}, Cosme \cite{Mor02}, DAMA \cite{Ber01},
Tokyo helioscope \cite{Ino02} and CAST \cite{Ari09} experiments. The areas of excluded values are located above the
corresponding lines} \label{fig4}
\end{figure}
The minimum of $\chi^{2}/n.d.f.=121.9/115$ corresponds to the
nonphysical value of the axion peak area $S_A=-14$ events. The upper
limit on the number of events within the peak was found via a
conventional approach: the dependence of $\chi^{2}$ on the peak area
$S_A$ was calculated for various values of $S_A$ while the rest of
the parameters were free. Then the appearance probability of the
given $\chi^{2}(S_A)$ value was found and the obtained function
$P(\chi^2(S_A))$ was normalized to unity for the $S_A\geq0$ region.
The upper limit appeared to be equal to $S_{lim}=31$ events for 90\%
c.l.

For the given rate of axion absorption $R_A$ the expected number of
registered 8.41 keV $\gamma$-quanta is:
\begin{equation}\label{slim}
    S_A=\varepsilon\cdot \eta\cdot N_{169Tm}\cdot T\cdot R_A=3.49\cdot 10^{23}\cdot R_A \leq
    S_{lim},
\end{equation}
where  $N_{169Tm}=1.09\cdot10^{21}$ - the number of $^{169}$Tm
nuclei, $T=6.61\cdot10^{6}$ s - time of measurement,
$\varepsilon=1.28\cdot10^{-2}$ - detection efficiency and
$\eta=3.79\cdot10^{-3}$ - internal conversion ratio.

The relation (\ref{slim}) obtained in the experiment ($R_A\leq8.8
\cdot$10$^{-23}$) limits the region of possible values of the
coupling constant $g_{A\gamma}$, $g_{AN}^0$, $g_{AN}^3$ and axion
mass $m_A$. In accordance with equations (\ref{rategamag0g3}),
(\ref{rategama}), (\ref{ratenumeric}) and on condition that
$(p_A/p_{\gamma})^3\cong1$ provided for $m_A < 2$ keV one can
obtain:
\begin{equation}\label{lim1}
g_{A\gamma}\cdot |(g_{AN}^0+g_{AN}^3| \leq 9.2\cdot 10^{-13}
\end{equation}
\begin{equation}\label{lim2}
g_{A\gamma}(\mbox{GeV}^{-1})\cdot m_A(\mbox{eV}) \leq 1.36\cdot
10^{-5}
\end{equation}
\begin{equation}\label{lim3}
m_A \leq 191 \mbox{ eV},
\end{equation}
all at 90\% c.l.

The limit (\ref{lim1}) is a model independent one on axion-photon and axion-nucleon couplings. The result (\ref{lim2}) (the
dimensionless quantity $g_{A\gamma}m_A\leq 1.36\cdot10^{-14}$) presented as a restriction on the range of possible values of
$g_{A\gamma}$ and $m_A$ (extracted from $g_{AN}$) allows us to compare our result with the results of experiments using the
conversion of axions to photon in the laboratory magnetic field or in the field of crystals \cite{Sik83}-\cite{Mor02}. The
comparisons are shown in Fig.\ref{fig4}. One can see that our results exclude a region of relatively large values of
$g_{A\gamma}$ and $m_{A}$. For $m_A \approx$ 1 keV the limit on the $g_{A\gamma}\approx$ 10$^{-8}$ GeV$^{-8}$ is only several
times lower than obtained for low axion masses. 

\begin{figure}
\includegraphics[width=9cm, height=10.5cm]{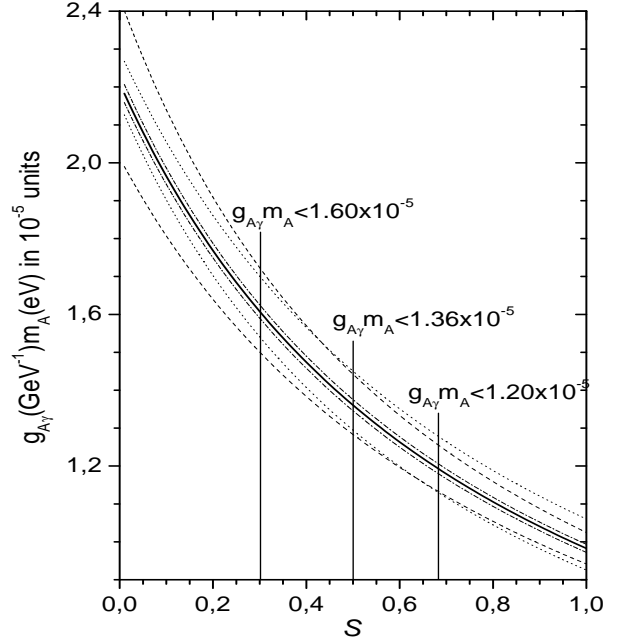}
\caption {The limit on $g_{A\gamma}(\mbox{GeV}^{-1})\cdot
m_A(\mbox{eV}^{-1})$ versus the value of $S$ parameter ($\beta$=1,
$\eta$=0.5, $z$=0.56, solid line ). The doted, dash-doted and dashed
lines correspond to the values of $\beta$, $\eta$ and $z$ changed on
$\pm$10\%, correspondingly.} \label{fig5}
\end{figure}

The limit on the hadronic axion mass ($m_A\leq$191 eV), obtained with the new method, is the strongest one up to date among the
results of  axion detection experiments based on the resonant absorption of monochromatic axions \cite{Krc98}-\cite{Bel08} or
axio-electric effect \cite{Lju04,Kek08}. The obtained limit depends on the exact values of the parameters $S$, $\beta$, $\eta$
and $z$. As mentioned above and in contrast with the 14.4 keV $^{57}$Fe solar axions \cite{Hax91}, the uncertainty of the
flavor-singlet axial-vector matrix element $S$ does not change the obtained constraints significantly: $m_A\leq$210 eV ($S$=0.3)
and $m_A\leq$180 eV ($S$=0.7). The changes of the nuclear-structure dependent terms $\beta$ from 0.5 to 1.5 and $\eta$ from -1
to +1 lead to the limits $m_A\leq$(223$\div$158) eV and $m_A\leq$(220$\div$180) eV, correspondingly. The usually accepted value
of $u$- and $d$ quark-mass ratio ($z$= 0.56) can vary in the range 0.3$\div$0.6 \cite{PDG08}, in this case our obtained results
become $m_A\leq$(163$\div$196) eV. The dependencies of the limits on $g_{A\gamma} m_A$ vs $S$ for the various values of $\beta$,
$\eta$ and $z$ are given in Fig.\ref{fig5}.

The sensitivity of the experiment depends on the total efficiency of
registration defined by the product $\eta\cdot\epsilon$ which is
$\approx 5\cdot 10^{-5}$ in our case. This value can be increased
significantly by introducing the Tm target inside the volume of a
scintillator.

\section{Conclusion}
The search for the resonant absorption of solar axions produced by
Primakoff conversion was performed. For that purpose we used the
low-background setup consisting of Si(Li)-detector, $^{169}$Tm
target, active and passive shielding. The obtained upper limit on
the values of the coupling constant and the axion mass is
$g_{A\gamma}(\mbox{GeV}^{-1})\cdot
m_A(\mbox{eV})\leq1.36\cdot10^{-5}$, which allowed us to set the
upper limit on hadronic axion mass $m_A\leq191$ eV (90\% C.L.)
($S$=0.5).

The work of E.V.Unzhakov was supported by a grant of
Saint-Petersburg Government, project no. 2.4/29-04/17.



\begin{thebibliography}{40}

 \bibitem{Pec77} R.D.~Peccei, H.R.~Quinn, Phys. Rev.~Lett. 38, 1440 (1977).

 \bibitem{Wei78} S.~Weinberg, Phys.~Rev.~Lett. 40, 223 (1978).

 \bibitem{Wil78} F.~Wilczek, Phys.~Rev.~Lett. 40, 279 (1978).

 \bibitem{Kim79} J.E.~Kim, Phys.~Rev.~Lett. 43, 103 (1979).

 \bibitem{Shi80} M.A.~Shifman, A.I.~Vainstein, and V.I.~Zakharov, Nucl.~Phys. B166, 493 (1980).

 \bibitem{Zhi80} A.R.~Zhitnitskii, Yad.~Fiz. 31, 497 (1980) (Sov.~J. ~Nucl.~Phys. 31, 260 (1980)).

 \bibitem{Din81} M.~Dine, F.~Fischler, and M.~Srednicki, Phys. Lett. B104, 199 (1981).

 \bibitem{Kap85} D.B.~Kaplan, Nucl.~Phys. B260, 215 (1985).


 \bibitem{Sik83} P.~Sikivie, Phys.~Rev.~Lett. 51, 1415 (1983), Phys.~Rev. D32, 2988 (1985).

 \bibitem{Kra87} L.~Krauss, et al., Phys.~Rev.~Lett. 55, 1797 (1987).

 \bibitem{Bib89} K.~van Bibber et al., Phys.~Rev. D39, 2089 (1989).

 \bibitem{Wue89} W.~Wuensch, et al., Phys.~Rev. D40, 3153 (1989).

 \bibitem{Hag98} C.~Hagmann, et al., Phys.~Rev.~Lett. 80, 2043 (1998).

 \bibitem{Muc01} M.~Muck, J.B.~Kycia, J.~Clarke, Appl.~Phys.~Lett. 78, 967 (2001).

\bibitem{Laz92} D.~Lazarus, et al., Phys.~Rev.~Lett. 69, 2333 (1992).

 \bibitem{Ino02} Y.~Inoue, et al., Phys.~Let. B536, 18 (2002), B668, 93 (2008).

\bibitem{Zio05} K.~Zioutas et al.,(CAST coll.), Phys.~Rev.~Lett. 94, 121301 (2005)

 \bibitem{Ari09} E. Arik et al., (CAST coll.), JCAP 0902.008 (2009), arXiv:0810.4482.

 \bibitem{Avi99} F.T.~Avignone et al., (Solax coll.) Nucl.~Phys., (Proc. Supll.) 72, 176 (1999).

 \bibitem{Ber01} R.~Bernabei, et al.,(DAMA coll.) Phys.~Lett. B515, 6 (2001).

 \bibitem{Mor02} A.~Morales et al., (Cosme coll.) Astropart.~Phys. 16, 325 (2002).

 \bibitem{Bru08} T.~Bruch, (CDMS coll.) arXiv:0811.4171 [astro-ph], (2008).


 \bibitem{Mor95} S.~Moriyama, Phys.~Rev.~Lett. 75, 3222 (1995).

 \bibitem{Krc98} M.~Kr\c{c}mar, et al., Phys.~Lett. B442, 38 (1998).

 \bibitem{Krc01} M.~Kr\c{c}mar, et al., Phys.~Rev. D64, 115016 (2001).

 \bibitem{Jak04} K.~Jakov\c{c}i\c{c}, et al., nuclex/0402016 (2004).

 \bibitem{Der05} A.V.~Derbin, et al., JETP Lett. 81, 365 (2005).

 \bibitem{Der07} A.V.~Derbin, et al., JETP Lett. 85, 12 (2007).

 \bibitem{Der07_A} A.V.~Derbin, et al., Bull.~Rus.~Acad.~Sci.~Phys. 71, 832 (2007).

 \bibitem{Nam07} T.~Namba, Phys.~Lett. B645, 398 (2007).

 \bibitem{Bel08} P.~Belli, et al., Nucl.~Phys. A806, 388 (2008).


 \bibitem{Raf08} S.~Hannestad, et al., arXiv:0706.9148, G.~Raffelt, et al., arXiv:0808.0814

 \bibitem{PDG08} C.~Amsler et al., (Particle Data Group) Phys.~Lett. B667, 1 (2008).

 \bibitem{Raf06}G.G.~Raffelt, arXiv:hep-ph/0611350


 \bibitem{Cre98} R.J.~Creswick, et al., Phys.~Lett. B427, 235 (1998).


 \bibitem{NDS08} Nuclear Data Sheets, A=169, 109 (2008).


 \bibitem{Don78} T.W.~Donnelly, et al., Phys.~Rev. D18, 1607 (1978).

 \bibitem{Avi88} F.T.~Avignone III, et al., Phys.~Rev. D 37, 618 (1988).

 \bibitem{Hax91} W.C.~Haxton and K. Y. Lee, Phys.~Rev.~Lett. 66, 2557 (1991).


 \bibitem{Sre85} M.~Srednicki, Nucl. Phys. B260, 689 (1985).

 \bibitem{Mat05} V.~Mateu and A.~Pich, J.~High Energy Phys. 10, 41 (2005).

 \bibitem{May89}R.~Mayle et al., Phys.~Lett. B219 (1989).


 \bibitem{Alt97} G.~Altarelli, et al., Phys.~Lett. B46, 337 (1997).

 \bibitem{Ada97} D.~Adams et al., Phys.~Rev. D56, 5330 (1997).


 \bibitem{Lju04} A.~Ljubicic, et al., Phys.~Lett. B599, 143 (2004).

 \bibitem{Kek08} D.~Kekez, et al., arXiv:0807.3482 (2008).

\end{thebibliography}
\end{document}